\documentstyle[12pt]{article}
\begin{document}

\begin{titlepage}

\begin{flushright}
OUTP-97-26P \\
gr-qc/9706007 \\
\end{flushright}
\begin{centering}
\vspace{.1in}
{\large {\bf Classicality, Matter-Antimatter Asymmetry,
and Quantum-Gravity Deformed Uncertainty Relations}
\footnote{This essay received 
``honorable mention'' from the Gravity Research Foundation, 1997 --- Ed.}} \\
\vspace{.2in}
{\bf 
Giovanni AMELINO-CAMELIA}

\vspace{.2in}

Theoretical Physics, University of Oxford,
1 Keble Rd., Oxford OX1 3NP, U.K. \\

\vspace{.3in}
\vspace{.1in}
{\bf Abstract}\\
\end{centering}

\vspace{.05in}

\baselineskip=12pt

{\small
Some of the recent work on quantum gravity has involved
modified uncertainty relations such that
the products of the uncertainties of certain pairs
of observables increase with time.
It is here observed that this type of modified uncertainty 
relations would lead to quantum decoherence, which
could explain the classical behavior of macroscopic systems, 
and CPT violation, which coud provide
the seed for the emergence of a matter-antimatter asymmetry.}

\end{titlepage}

\newpage

\baselineskip 12pt plus .5pt minus .5pt

\pagenumbering{arabic}
\setcounter{page}{1}
\pagestyle{plain} 

\section{Introduction}

Since a compelling theory encompassing gravity and quantum mechanics
has not yet emerged (although substantial progress has been made
along certain lines of research,
most notably in critical string theory~\cite{gsw,dbrane} and
in  canonical quantum gravity~\cite{asht}),
the physics community continues to be tempted
by the possibility
that such a theory
might also provide the resolution to other long-standing puzzles
confronting our present description of Nature.
In particular, the possibility that ``quantum gravity"
might solve the problems associated to the quantum-mechanical
analysis of macroscopic
systems (which we find to behave classically in apparent
violation of quantum mechanics)
has been considered by several authors.
In fact, the emergence of classicality could be attributed
to gravitational effects that induce quantum decoherence
(see, {\it e.g.}, \cite{qgclass}).

This Essay is devoted to  certain quantum-gravity modified
uncertainty relations that have been
recently discussed both in heuristic analyses of the measurability 
of distances in quantum gravity
and in the context of quantum-gravity theories based
on noncritical strings or quantum Poincar\`e symmetries.
It will be here emphasized that such modified uncertainty relations
could lead to quantum decoherence, with the
above-mentioned implications for the understanding of classicality,
and CPT violation, which could provide the seed for
the emergence of a matter-antimatter asymmetry.

The ``understanding'' of the
observed matter-antimatter asymmetry is another
long-standing problem of theoretical physics.
The theories presently used to describe particle physics
do not naturally accommodate an asymmetry of the observed size.
A substantial effort has gone into studies
(see, {\it e.g.}, Ref.~\cite{baryo}) that exploit CP violation 
to generate the asymmetry.
CPT violation is not usually considered, since the
relativistic field theories describing (non-gravitational)
particle physics are necessarily CPT invariant;
however, it is well understood (see, {\it e.g.}, Ref.~\cite{cptbaryo})
that CPT violation would provide a rather natural seed
for the emergence of a matter-antimatter asymmetry.

In presenting the line of argument here advocated,
an attempt is made in the following of proceeding 
through intuitive observations concerning the relevant structures, 
rather than providing detailed/technical analyses,
which would go beyond the scope of this Essay.
Details concerning the quantum-gravity modified uncertainty relations
discussed in the next section and the decoherence
mechanism discussed in Sec.~3
can be found in 
Refs.~[7-10].
A detailed study of the CPT-violation mechanism
discussed in Sec.~3 together with its implications
for the emergence of a matter-antimatter asymmetry
will be provided in Ref.~\cite{gacprep}.

\section{Quantum-gravity modified uncertainty relations}

The expectation that Heisenberg's uncertainty relations
would have to be modified in order to accommodate (quantum)
gravitational effects has been often expressed.
This can be justified on several grounds,
with the non-covariance of the relations
perhaps providing the most intuitive argument.
The presence of a natural length scale, the Planck length $l_p$,
in the quantum-gravitational context is
rather naturally expected to play a role in the modifications.
In particular, a common 
hypothesis (see, {\it e.g.}, Ref.~\cite{padma,venezkonish})
is that the
Planck length would set an absolute lower bound on the measurability
of distances. Such a bound would provide a clear departure from
ordinary quantum mechanics, in which any observable
can be measured (in principle)
with complete accuracy (at the price of loosing all
information on a conjugate observable).
A rather ``conservative" scenario that accommodates $l_p$ as
minimum uncertainty
is provided by the following modification of
Heisenberg's space-momentum uncertainty relation
(henceforth
$\hbar \! \sim \! c \! \sim \! 1 \sim \! 2$)
\begin{eqnarray}
 \delta x \, \delta p \ge  1
+ {l_p^2} \, \delta p^2
~. \label{veneup}
\end{eqnarray}
Relations of type (\ref{veneup}) have been quite extensively
investigated, especially in light of their relevance
for critical string theory, in which
evidence in support of the relation (\ref{veneup}), but
with $l_p$ replaced by the string length $l_s$,
is found in the analysis of high-energy
string scattering~\cite{venezkonish}.

Even within the framework of critical string theory,
more drastic modifications of quantum mechanics have been
considered; in particular,
the possibility of non-trivial space-time
uncertainty relations of the type 
\begin{eqnarray}
 \delta x \delta t \ge l_s^2
~ \label{dxdtls}
\end{eqnarray}
has been studied in detail~\cite{yoneya}.

The issue of modified uncertainty relations
continues to be quite central in critical
string theory, and the recent developments~\cite{dbrane} in the
understanding of non-perturbative structures
have already been exploited from this point of view.
Both (\ref{veneup}) and (\ref{dxdtls}) have been reanalyzed in recent 
studies~\cite{dbrscatt,yoneyanew}
of the soliton-like structures known as
Dirichlet p-branes~\cite{dbrane},
and, interestingly, evidence has been found~\cite{dbrscatt}
in support of the idea that ``D-particles"  (Dirichlet 0-branes)
could probe the structure of space-time down
to scales shorter than the string length,
raising the possibility that (\ref{veneup}) might have to be modified.

In parallel with these developments in the literature on
critical string theories and conventional quantum-gravity approaches,
there have been studies of modified uncertainty relations
in the context of models of
quantum gravity that support quantum decoherence,
such as certain non-critical string theories~\cite{emn}
and certain theories based
on quantum Poincar\`e symmetries~\cite{review}.
It is this side of the debate on quantum-gravity
modified uncertainty relations
that is relevant for the mechanisms advocated in this Essay.
The emergent general expectation is that
decoherence effects should cause the
uncertainties characterizing a measurement procedure
to grow with the time required by the procedure.
In particular, as discussed in Ref.~\cite{gacmpla}, 
the fact that gravitational effects prevent one from relying
on {\it classical} agents for the
measurement procedure\footnote{\tenrm
\baselineskip = 11pt
 As emphasized in Ref.~\cite{gacmpla},
``classical'' ({\it i.e.} infinitely massive) devices,
whose position and velocity are both completely determined,
would lead to inconsistencies associated with the formation of
horizons.
It is worth noticing that the infinite-mass limit
is crucial for ordinary quantum mechanics, which after all is a theory
describing the results of experiments in which classical devices
observe properties of a microscopic system.
The fact that the infinite-mass limit should not be viable in quantum gravity,
besides having the implications for the measurability of distances
discussed in Ref.~\cite{gacmpla},
can be expected to affect also the measurability of the 
gravitational field.
In the famous analysis by Bhor and Rosenfeld~\cite{rose},
which established that 
the electromagnetic field is measurable with complete accuracy
in ordinary quantum mechanics, a crucial role is played by
classical probes, {\it i.e.} ideal probes
whose ratio of electric charge versus inertial mass
can be taken to zero.
Although we don't have a fully satisfactory 
quantum gravity, we can expect that the gravitational field
be not measurable with complete accuracy,
since the equivalence principle
demands that for all probes the
ratio of gravitational charge versus inertial mass be 1.
Some work relevant to this line of argument can be found in 
Ref.~\cite{bergstac}.},
leads to the following bound for the measurability
of a distance $L$:
\begin{eqnarray}
min \left[ \delta L \right] \sim
\sqrt{ {\eta T}} \sim
\sqrt{ {\eta L}}
~,
\label{qgboundgac}
\end{eqnarray}
where $\eta$ is a (dimensionful) parameter characterizing~\cite{gacmpla}
the spatial extension of the devices ({\it e.g.}, clocks) 
used in the measurement, $T$ is the 
time needed to complete the procedure of measuring $L$,
and the right-hand side takes into account
the fact that $T$ is typically~\cite{gacmpla}
proportional to $L$.
A candidate modified space-momentum
uncertainty relation that 
leads to the bound (\ref{qgboundgac}) 
is given by~\cite{gacxt}
\begin{eqnarray}
 \delta x \, \delta p  \ge 1 + \eta 
\, T \, \delta p^2
~, \label{modup}
\end{eqnarray}
but the space-time
uncertainty relation
\begin{eqnarray}
\delta x \, \delta t \ge {\eta x}  \sim {\eta t} 
 \label{newonegeneral}
\end{eqnarray}
would lead to the same bound as a result of its implications
for the measurement of distances~\cite{gacxt}.

Besides the measurement analyses reported in Refs.~\cite{gacmpla,gacxt},
and the related studies~\cite{karo}, evidence for 
bounds of type (\ref{qgboundgac}) have been found directly in certain
approaches to quantum gravity.
In particular,
an uncertainty relation of the form (\ref{modup})
appears~\cite{aemn} to characterize 
Liouville (non-critical) string 
theories, upon the interpretation of the Liouville mode
as the target time~\cite{emn}, 
while an uncertainty relation of the type (\ref{newonegeneral})
has emerged in the context
of studies of the quantum $\kappa$-Poincar\`e group~\cite{review},
which could play a role~\cite{gacxt} in the
description
of the nature of geometry at distances not much larger than the Planck length.

\section{Decoherence and CPT violation}

While in the previous section I emphasized that the
measurability
bound (\ref{qgboundgac}),
and uncertainty relations such as
(\ref{modup}) and (\ref{newonegeneral}),
could naturally emerge in theories supporting quantum decoherence, 
I now want to clarify that in turn it is also true that quantum decoherence
is induced in formalisms
involving this type of relations.
This is discussed at length in Ref.~\cite{karo};
however, it is worth illustrating it here at an intuitive level.
Let us take for example the uncertainty relation (\ref{modup}).
A system prepared as a pure state completely localized at $x \! = \! x_0$
would accordingly evolve into a (mixed) state
with $x \! = \! x_0 \pm \sqrt{\eta t}$.
Quantum coherence would therefore be destroyed (at least in the
sense of ordinary quantum mechanics).

In order to show that 
relations such as
(\ref{qgboundgac}),
(\ref{modup}), and (\ref{newonegeneral})
naturally lead to CPT violation it is useful to investigate
field theories in accordingly deformed phase spaces.
A detailed analysis of this issue
(within the limitations set by the present
poor development of such field theories~\cite{noncommFT}) 
will be reported in Ref.~\cite{gacprep}.
[Related results can also be found in the studies
of the relation between
decoherence and CPT violation reported in
the Refs.~\cite{emncpt,peskin}.]
Consistently with the general tone of this Essay,
here I try to present an intuitive argument
illustrating the emergence of CPT violation
when structures of the type discussed in
the preceding section are present.
This argument requires that alongside the space-momentum
uncertainty relation (\ref{modup}),
the corresponding
time-energy uncertainty relation
\begin{eqnarray}
\delta t \, \delta \! E \ge
1 + \eta t \, \delta \! E^2
~ \label{teupgacng}
\end{eqnarray}
should also hold. This would be in analogy with the situation in ordinary
quantum mechanics, in which a
time-energy uncertainty relation does hold alongside Heisenberg's
space-momentum uncertainty relation.\footnote{\tenrm
\baselineskip = 11pt
However, whereas Heisenberg's operatorial space-momentum
uncertainty relation is rooted deep into the formalism,
the time-energy uncertainty relation comes in at a somewhat less
fundamental level, and cannot be interpreted as an operatorial relation.}
Assuming (\ref{teupgacng}) one would then expect the
lifetime $t$ of an unstable state to be related
to the average lifetime $\tau$ (time uncertainty)
of an ensemble of such states
and the corresponding level width $\Gamma$ (energy uncertainty) by
\begin{eqnarray}
\Gamma \sim
{1 \over \tau} + {\eta \over \tau^3} t
~. \label{gammatau}
\end{eqnarray}
This equation is formally equivalent to one that could be obtained
by introducing time-dependent decay rates, {\it i.e.}
time-dependent couplings. It is therefore not surprising that
a formalism supporting (\ref{gammatau}) would violate CPT invariance.
(Naively antiparticles propagate backward in time, and therefore
time-dependent couplings can affect particles and antiparticles differently.)

The idea of quantum-gravity induced CPT violation
is actually not so foreign to the quantum-gravity literature.
In particular, the type of non-locality advocated in Ref.~\cite{alu}
and the description of the space coordinates as
macroscopic variables of a statistical system
discussed in Ref.~\cite{farmat}
provide other frameworks in which 
CPT violation could emerge naturally.

\section{Closing remarks}
This author has here failed to resist the common temptation
of hoping that quantum gravity
might provide the resolution to several long-standing puzzles
confronting our present description of Nature. From this point of 
view some of the structures discussed
in Sec.2 have been found to be quite promising,
most notably for decoherence (classicality) and CPT violation
(matter-antimatter asymmetry),
and this should encourage further exploration of the related physics.
However, the approaches to quantum gravity in which evidence
of such structures has emerged, such as
noncritical string theory and field theory with $\kappa$-deformed
Poincar\`e symmetries,
are still very poorly developed, and it cannot be excluded that
the complete understanding of such approaches would not support
any of the relations
(\ref{qgboundgac}),
(\ref{modup}), and (\ref{newonegeneral}).

Critical string theory, while being
the quantum-gravity approach whose development
has been most successful,
does not appear to accommodate naturally
an intrinsic microscopic mechanism for decoherence
or a scenario for the emergence of a matter-antimatter asymmetry.
In general, critical string theory appears to provide a rather conservative
(while blessed by the emergence of remarkable mathematical structures)
modification of our understanding of nature.
This is reflected in the ``conservative'' 
deformation (\ref{veneup}) of Heisenberg's uncertainty
relation and in the nature of the
recent stringy understanding~\cite{stringBH} of the
black hole ``paradox".

\newpage

\baselineskip 12pt plus .5pt minus .5pt

\end{document}